\documentclass[twocolumn]{jpsj2}

\newcommand{\msr}{$\mu$SR}
\newcommand{\lfaof}{LaFeAsO$_{1-x}$F$_x$}

\newcommand{\bkfa}{Ba$_{1-x}$K$_x$Fe$_2$As$_2$}

\title{
Full Gap Superconductivity in Ba$_{0.6}$K$_{0.4}$Fe$_2$As$_2$ Probed by Muon Spin Rotation
}

\author{
Masatoshi \textsc{Hiraishi}$^{1}$,
Ryosuke \textsc{Kadono}$^{1,2}$\thanks{E-mail address : ryosuke.kadono@
kek.jp},
Soshi \textsc{Takeshita}$^{2}$,\\
Masanori \textsc{Miyazaki}$^{1}$,
Akihiro \textsc{Koda}$^{1,2}$,
Hirotaka \textsc{Okabe}$^{3}$
and
Jun \textsc{Akimitsu}$^{3}$
}

\inst{
$^{1}$Department of Materials Structure Science, School of High Energy
Accelerator Science, \\The Graduate University for Advanced Studies,
1-1 Oho, Tsukuba, Ibaraki 305-0801\\
$^{2}$Muon Science Laboratory, Institute of Materials Structure Science, High Energy Accelerator
Research Organization (KEK), 1-1 Oho, Tsukuba, Ibaraki 305-0801\\
$^{3}$Department of Physics and Mathematics, Aoyama-Gakuin University, 
Fuchinobe, Sagamihara, Kanagawa, 229-8558\\
}

\abst{Superfluid density ($n_s$) in the mixed state of an iron pnictide superconductor \bkfa\ is determined by muon spin rotation for a sample with optimal doping ($x=0.4$). The temperature dependence of $n_s$ is perfectly reproduced by the conventional BCS model for $s$-wave paring,  where the order parameter can be either a single-gap with $\Delta=8.35(6)$ meV [$2\Delta/k_BT_c=5.09(4)$], or double-gap structure with $\Delta_1=12$ meV (fixed) [$2\Delta_1/k_BT_c=7.3$] and $\Delta_2=6.8(3)$ meV [$2\Delta_2/k_BT_c=4.1(2)$]. The latter is consistent with the recent result of angle-resolved photo-emssion spectroscopy. The large gap parameters ($2\Delta/k_BT_c$) indicate extremely strong coupling of carriers to bosons that mediate the Cooper pairing.
}
\kword{oxypnictide superconductor, order parameter,  muon spin rotation}

\begin{document}
\maketitle


The recent discovery of high-$T_c$ superconductivity in \lfaof\ (LFAO-F, $T_{\rm c}\simeq 26$~K) and of related compounds \cite{R_Kamihara,HCChen:08,ZARen:08,ZARen:08-2,Kito:08,GFChen:08,Sefat:08,Sefat:08-2,Qi:08,Matsuishi:08,Rotter:08,Sasmal:08,GFChen:08-2,GWu:08}  has attracted broad attention regarding the mechanism of superconductivity on Fe$_2$As$_2$ layers that show some apparent similarities with CuO$_2$ layers.  They have a common feature that the superconductivity occurs 
upon carrier doping to the transition metal oxide/pnictide layers that exhibit instability towards magnetic order at lower temperatures.  In the case of LFAO-F and its family compounds, {\sl electrons} are introduced by the substitution of O$^{2-}$ with F$^{-}$ in the La$_2$O$_2$ layers\cite{R_Kamihara,HCChen:08,ZARen:08,GFChen:08}. Moreover, very recent developments demonstrate increasing variety in the methods of electron doping such as oxygen depletion \cite{Kito:08,ZARen:08-2} or Co substitution for Fe\cite{Sefat:08,Sefat:08-2,Qi:08,Matsuishi:08}.
In contrast, {\sl hole} doping is attained  in $A$Fe$_2$As$_2$ ($A=$ Ba, Sr, Ca) by the substitution of divalent cations ($A^{2+}$) with alkali metals ($B^{+}$) \cite{Rotter:08,Sasmal:08,GFChen:08-2,GWu:08}. This situation provides an excellent opportunity to examine the ``electron-hole symmetry" regarding the superconducting properties of iron pnictides, which is also one of the major issues in cuprates.

Microscopic evidence shows that the parent compound BaFe$_2$As$_2$ exhibits magnetic order (SDW) below 140 K which is accompanied by a structural phase transition \cite{Rotter:08-2,Aczel:08}. The situation is common to Sr$_{1-x}$K$_x$Fe$_2$As$_2$ \cite{Zhao:08}, and it suggests that the electronic ground state of Fe$_2$As$_2$ layers in the parent compound is quite similar to that in LaFeAsO.  However,  the doping phase diagram is markedly different from LFAO-F, as it exhibits superconductivity over a wide range of hole content $p=x/2$ (per FeAs chemical formula) from 0.05 to 0.5 that far exceeds LFAO-F (i.e., $0.06\le x\le 0.2$)\cite{Sasmal:08}.  The phase diagram is also characterized by a bell-shaped variation of $T_c$ against $x$, where the maximal $T_c\simeq38$ K is attained near $x\sim0.4$ ($p\sim0.2$).  Considering the manifold nature of electronic band structure of Fe$_2$As$_2$ layers suggested by theories\cite{Mazin:08,Kuroki:08,Nagai:08}, this difference may be attributed to that of the bands relevant to the doped carriers.  Here, we report our \msr\ study on \bkfa\ (BKFA) for samples with $x=0.1$ ($T_c<2$ K) and 0.4 ($T_c\simeq38$ K). The result suggests that, while the magnetic ground state for $x=0.1$ is mostly identical with that of other pristine iron pnictides, the superconducting property in the optimally-doped sample is characterized by gap parameters that are much greater than  those of LFAO-F\cite{Luetkens:08,Carlo:08,Takeshita:08} and CaFe$_{1-x}$Co$_x$AsF (CFCAF, another subclass of electron-doped iron pnictide superconductor)\cite{Takeshita:08-2}. 

Polycrystalline samples of \bkfa\ with nominal compositions of $x=0.1$ and 0.4 were prepared by a solid state reaction using Ba (99\%, Furuya Metal), K (99\%, Kojundo-Kagaku), Fe (99.9\%, Kojundo-Kagaku) and As (99.9999\%, Furuchi Chemical) as starting materials, where details of the preparation process are described in an earlier report\cite{GWu:08}. Concerning the K-doped samples, reproducibility of synthesis was improved by using barium and potassium 
arsenides as precursors.  The purity of as-grown samples (sintered slabs with a dimension of $\sim10 \times 12$--$15\times 1$ mm$^3$, net weight of $\sim0.3$ g) was examined by powder X-ray diffraction using a diffractometer with a graphite monochromator (MultiFlex, Rigaku). All the observed diffraction peaks in the sample with $x=0.4$ was perfectly reproduced by those of single phase compound\cite{Rotter:08}, while a minor unknown impurity phase ($\sim$2\%) was observed for $x=0.1$. Conventional $\mu$SR measurements were performed using the LAMPF spectrometer installed on the M20 beamline of TRIUMF, Canada. During the measurement under a zero field (ZF), residual magnetic field at the sample position was reduced below $10^{-6}$~T with the initial muon spin direction parallel to the muon beam direction [$\vec{P}_\mu(0)\parallel \hat{z}$].  
Time-dependent muon polarization [$G_z(t)=\hat{z}\cdot \vec{P}_\mu(t)$] was monitored by measuring decay-positron asymmetry along the $\hat{z}$-axis.  Transverse field (TF) condition was realized by
rotating the initial muon polarization so that $\vec{P}_\mu(0)\parallel
\hat{x}$, where the asymmetry was monitored along the $\hat{x}$-axis to obtain $G_x(t)=\hat{x}\cdot \vec{P}_\mu(t)$.  All the measurements under a magnetic field were made by cooling the sample to the target
temperature after the field equilibrated.

\begin{figure}[tp]
\begin{center}
\includegraphics[width=0.45\textwidth,clip]{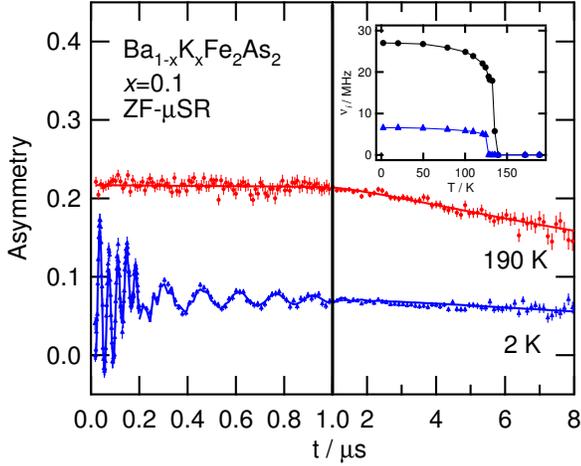}
\caption{(Color online)
ZF-$\mu$SR time spectra observed at 190 K and 2 K in \bkfa\ sample with $x=0.1$ ($T_c<2$K), where a spontaneous muon precession signal (mainly consisting of two frequencies) is clearly seen. Inset shows temperature dependence of the precession frequency. }
\label{bfa-tsp}
\end{center}
\end{figure}

Figure \ref{bfa-tsp} shows the \msr\ spectra obtained upon muon implantation to the sample with $x=0.1$ under zero external field, where one can readily identify an oscillatory signal with multiple frequency components in the spectrum at 2 K.  The Fourier transform of the spectrum indicates that there are actually two components, one approaching to $\nu_1=27.0(1)$ MHz and another to $\nu_2=6.53(2)$ MHz [at 2 K, corresponding to  internal fields of 0.20(1) T and 0.0482(1) T, respectively].  This is consistent with earlier result of \msr\ study in \bkfa, where they observed signals of 28.8 MHz and 7 MHz in a parent compound ($x=0$)\cite{Aczel:08}.  While the quality of their K-doped sample ($x=0.45$) seems to have some problem regarding homogeneity (see below), its magnetic phase exhibits a tendency that  $\nu_1$ is slightly reduced ($\sim26$--27 MHz).  
This is quite in line with the slight reduction of $\nu_1$ observed for our sample with $x=0.1$, and naturally understood as a result of enhanced itinerant character for $d$ electrons in K-doped compounds.  The magnitude of internal field probed by \msr\ is close to that observed in LFAO ($\sim23$ MHz and 3 MHz) \cite{Carlo:08,Klauss:08}, and thereby it suggests that the high frequency component corresponds to the signal from muons situated on the Fe$_2$As$_2$ layers while another coming from those located near the cation sites. The onset temperature for the high frequency component is close to 140 K, which is also consistent with earlier reports \cite{Rotter:08-2,Aczel:08}.

\begin{figure}[tp]
\begin{center}
\includegraphics[width=0.45\textwidth,clip]{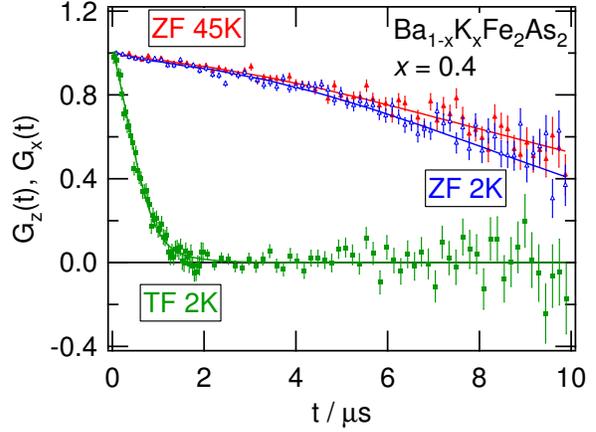}
\caption{(Color online)
 ZF-$\mu$SR time spectra observed at 45 K and 2 K in \bkfa\ sample with $x=0.4$ ($T_c\simeq38$ K). No trace of magnetic phase is observed. The spectra under a transverse field (TF) at 2 K is displayed on a rotating reference frame to extract the envelop function. Solid curves are fits using a Gaussian relaxation function described in the text. }
\label{bkfa-tsp}
\end{center}
\end{figure}

\begin{figure}[tp]
\begin{center}
\includegraphics[width=0.45\textwidth,clip]{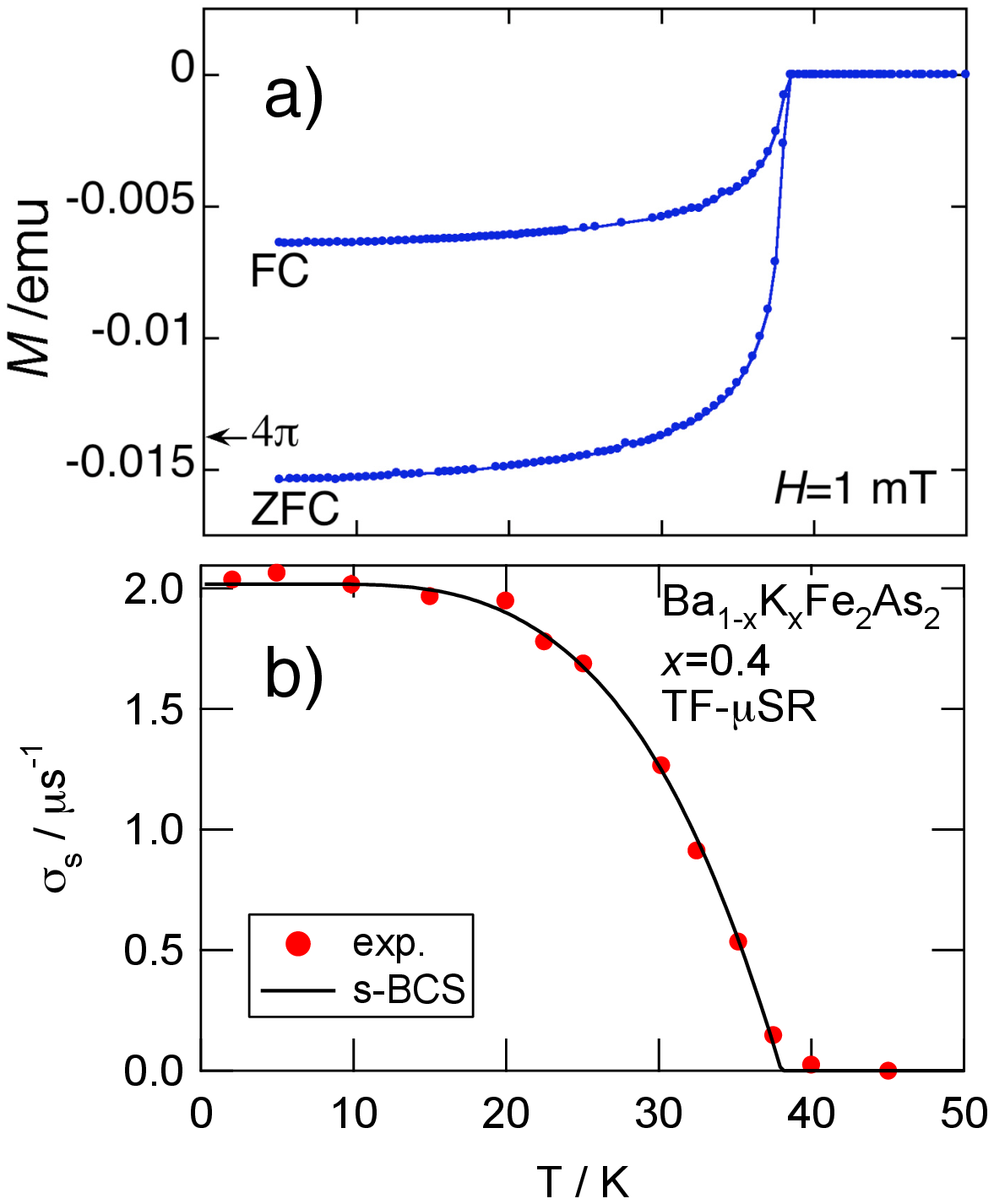}
\caption{(Color online)
a) Magnetization of \bkfa\ measured on the \msr\ sample with $x=0.4$, where data were obtained after cooling under an external field (FC) or zero field (ZFC). The total weight of the sample is 0.1 g, and the magnetization corresponding to $4\pi$ is obtained by using the structural parameters reported in Ref.\cite{Rotter:08}.
b) Temperature dependence of Gaussian linwidth $\sigma_{\rm s}=\sqrt{2}\delta_{\rm s}$  determined by TF-\msr\ measurement with $H=0.1$ T. Solid curve is a fit using weak coupling BCS model (see text for detail).}
\label{bkfa-chi}
\end{center}
\end{figure}

On the other hand, it is inferred from ZF-\msr\ spectra in Fig.~\ref{bkfa-tsp} that no trace of magnetism is found in the sample with $x=0.4$. This is in marked contrast with the earlier report on \msr\ measurements in a sample with $x=0.45$ where a magnetic phase seems to dominate over a large volume fraction ($\sim80$\%)\cite{Aczel:08}.  Concerning the bulk superconducting property, the quality of our specimen can be assessed by looking into magnetization data, which is shown in Fig.~\ref{bkfa-chi}a.  The sharp onset as well as a large Meissener fraction ($\ge4\pi$ for ZFC) endorses  excellent quality of the present specimen.

The \msr\ time spectra under a transverse field ($\mu_0H=0.1$ T) is shown in Fig.~\ref{bkfa-tsp}, where the envelop of the damping oscillation is extracted (by displaying on a rotating reference frame).  It exhibits depolarization towards zero, indicating that the entire sample falls into the flux line lattice state to exert strongly inhomogeneous internal field [$B({\bf r})$] to implanted muons. The lineshape is well represented by a Gaussian damping, and the analysis is made by curve fit using the following model function,
\begin{equation}
\label{E_TFG2}
 G_x(t)=\exp\left[-\frac{1}{2}(\delta^2_{\rm N}+\sigma^2_{\rm s})t^2\right]\cos{(2\pi f_{\rm s} t+\phi)},
\end{equation}
where $\sigma_{\rm s}=\gamma_\mu\langle(B({\bf r})-B_0)^2\rangle^{1/2}$ with $B_0\simeq \mu_0 H$ being the mean value of the local field $B({\bf r})$ \cite{Brandt:88}, $2\pi f_s\simeq\gamma_\mu B_0$ (with $\gamma_\mu=2\pi\times135.53$ MHz/T), and $\delta_{\rm N}$ is the depolarization due to random local fields from nuclear magnetic moments.  For the extraction of $\sigma_{\rm s}$, $\delta_{\rm N}$ is determined by curve fits above $T_c$ and then subtracted from the total linewidth obtained for the spectra below $T_c$.

The deduced linewidth, $\sigma_{\rm s}$, is plotted against temperature in Fig.~\ref{bkfa-chi}b, where $\sigma_{\rm s}$ is the quantity proportional to the superfluid density $n_{\rm s}$,
\begin{equation}
\sigma_{\rm s}\propto\frac{1}{\lambda^2}=\frac{n_{\rm s}e^2}{m^*c^2}\:,
\end{equation}
where $\lambda$ is the effective London penetration depth and $m^*$ is the effective mass of the superconducting carriers.
  Compared with the results of \msr\ studies in LFAO-F \cite{Luetkens:08,Carlo:08,Takeshita:08} and that recently obtained for CFCAF\cite{Takeshita:08-2}, it is noticeable that $\sigma_{\rm s}$ rises relatively sharply just below $T_c$, and becomes mostly independent of temperature below 15 K ($\simeq0.4T_c$).  A curve fit using the power law, $\sigma_{\rm s} =\sigma_0 [1-(T/T_{\rm c})^\beta]$, yields $\beta= 4.08(5)$, which is perfectly in line with the prediction of conventional BCS model for $s$-wave pairing.  The gap parameter is obtained by a fit using the weak coupling BCS model to yield $\Delta=8.35(6)$ meV and corresponding ratio 
$2\Delta/k_BT_c=5.09(4)$.  While these values are consistent with the superconducting order parameter of the isotropic $s$-wave pairing, the gap parameter far exceeds the prediction of weak-coupling BCS theory ($2\Delta/k_BT_c=3.53$), and thereby suggesting a very strong coupling of superconducting carriers to some bosonic excitations.

It is inferred from recent angle-resolved photoemission spectroscopy (ARPES) study on BKFA with the same potassium content that the magnitude of superconducting gap depends on the Fermi surfaces \cite{Ding:08}.  They report $\Delta_1\sim12$ meV on small Fermi surfaces and $\Delta_2\sim6$ meV on the large one.  We examined the consistency of our data with the ARPES result by employing a phenomenological double-gap model with $s$-wave symmetry \cite{Bouquet,Ohishi},
\begin{eqnarray}
\sigma_{\rm s}(T)&=&\sigma(0)-w{\cdot}\delta\sigma(\Delta_1,T)-(1-w){\cdot}\delta\sigma(\Delta_2,T),  \nonumber \\
\delta\sigma(\Delta,T)&=& \frac{2\sigma(0)}{k_{\rm B}T}\int_0^{\infty}f(\varepsilon,T){\cdot}[1-f(\varepsilon,T)]d\varepsilon, \nonumber \\
f(\varepsilon,T) &=& \left(1+e^{\sqrt{\varepsilon^2+\Delta(T)^2}/k_{\rm B}T}\right)^{-1},\nonumber 
\end{eqnarray}
where $\Delta_i$ ($i$ = 1 and 2) are the energy gap at $T=0$, $w$ is the relative weight for $i=1$, $k_{\rm B}$ is the Boltzmann constant, $f(\varepsilon,T)$ is the Fermi distribution function, and $\Delta(T)$ is the standard BCS gap energy. The curve fit assuming a common $T_c=38$ K and a large gap fixed to 12 meV ($2\Delta_1/k_BT_c=7.3$) perfectly reproduces data in Fig.~\ref{bkfa-chi}b with $\Delta_2=6.8(3)$ meV [$2\Delta_2/k_BT_c=4.1(2)$] and the relative weight $w=0.30(3)$, where the obtained curve is virtually identical with that for the single gap on Fig.~\ref{bkfa-chi}b.  This, while endorsing the credibility of our data in terms of temperature dependence of $\sigma_s$, indicates that the quasiparticle excitation spectrum associated with multi-gap superconductivity tends to be merged into that of the single gap when the small gap has a large value for $2\Delta/k_BT_c$.  

It has been shown in our previous \msr\ studies on LFAO-F \cite{Takeshita:08} and CFCAF\cite{Takeshita:08-2} that the temperature dependence of $\sigma_{\rm s}$ in these compounds (with a doping range near the boundary between magnetic and superconducting phases) can be reproduced by the above mentioned double-gap model, where the gap parameters are considerably smaller than those in BKFA [e.g., $2\Delta_1/k_BT_c=2.6(3)$, $2\Delta_2/k_BT_c=1.1(3)$ in LFAO-F with $x=0.06$, and $2\Delta_1/k_BT_c=3.7(8)$, $2\Delta_2/k_BT_c=1.1(7)$ in CFCAF with $x=0.075$].
The double-gap feature revealed by ARPES supports a view that  superconductivity occurs on complex Fermi surfaces consisting of many bands (at least five Fe $d$ bands) that would give rise to certain intricacy \cite{Mazin:08,Kuroki:08,Nagai:08}. Apart from the validity of applying the double-gap model to electron-doped iron pnictides, these figures suggest that the hole doping may occur in the bands different from those for electron doping, where the characteristic energy of the Cooper pairing may differ among those bands.  

\begin{figure}[tp]
\begin{center}
\includegraphics[width=0.45\textwidth,clip]{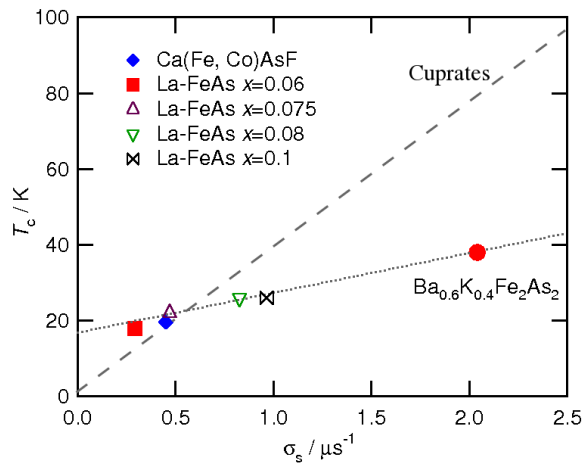}
\caption{(Color online) A plot of superconducting transition temperature 
vs muon spin relaxation rate $\sigma_s(T\rightarrow0)$, where a solid circle shows
the present data. Other points are quoted selectively for relevant iron pnictides (LFAO-F\cite{Carlo:08,Luetkens:08,Takeshita:08}, CFCAF\cite{Takeshita:08-2}).  The empirical relation is indicated by a dashed line\cite{Uemura:04}, and the dotted line
connecting points for iron pnictides is only meant for a guide for eye.}
\label{uplot}
\end{center}
\end{figure}

Finally, we point out that the superfluid density in the optimally doped \bkfa\ does not satisfy the empirical linear relation with $T_c$ observed for underdoped cuprates.\cite{Uemura:04} As shown in Fig.~\ref{uplot}, the corresponding muon spin relaxation rate [$\sigma_s(0)=2.02(1)$ $\mu$s$^{-1}$, which yields the magnetic penetration depth $\lambda=155(1)$ nm] is more than twice as large as what is expected for compounds with $T_c=38$ K from the empirical line shown by a dashed line.  
The data for other iron pnictides are also plotted for comparison, in which we omitted those for the compounds containing rare earth elements (Ce, Nd, Sm,...) because of the general ambiguity anticipated for extracting $\sigma_s$ under strong additional depolarization due to the magnetism of rare earth ions.  They show a tendency that $\sigma_s$ (and accordingly $n_{\rm s}$) is independent of $T_c$.  This is qualitatively different from underdoped cuprates, and rather close to the behavior predicted by the conventional BCS theory where the condensation energy does not depend 
on $n_{\rm s}$.

In summary, we have shown in a hole-doped iron pnictide, \bkfa, that the superconducting 
order parameter is characterized by a strong coupling to paring bosons, as inferred from large gap parameters ($2\Delta_i/k_BT_c\gg3.53$). The temperature dependence of the superfluid density, $n_{\rm s}(T)$,  determined by \msr\ is perfectly in line with that predicted by the conventional BCS model with fully gapped $s$-wave pairing.  A detailed analysis with phenomenological double-gap model indicates that $n_{\rm s}(T)$ is also consistent with the presence of double-gap, although the large gap parameters make it difficult to determine the multitude of energy gap solely from $n_{\rm s}(T)$.

\par

We would like to thank the TRIUMF staff for their technical support
during the $\mu$SR experiment. This work was partially supported by the
KEK-MSL Inter-University Program for Oversea Muon Facilities and by a
Grant-in-Aid for Creative Scientific Research on Priority Areas from the
Ministry of Education, Culture, Sports, Science and Technology, Japan.


\end{document}